\documentclass{article}
\usepackage{amssymb}

\usepackage{graphicx}
\usepackage{amsmath}

\newtheorem{theorem}{Theorem}

\newtheorem{definition}[theorem]{Definition}
\newtheorem{example}[theorem]{Example}
\newtheorem{exercise}[theorem]{Exercise}
\newtheorem{lemma}[theorem]{Lemma}
\newtheorem{notation}[theorem]{Notation}

\newtheorem{remark}[theorem]{Remark}

\newenvironment{proof}[1][Proof]{\textbf{#1.} }{\ \rule{0.5em}{0.5em}}

\begin{document}

\title{\textbf{A realistic and deterministic description of a quantum system}}
\author{\textbf{Antonio Cassa} \\
antonio.leonardo.cassa@gmail.com}
\maketitle

\begin{abstract}
From the analysis of the measurement process we make the hypothesis that we
have to add to the quantum state $[\psi ]$ a label $z$ and a special
function $\alpha $ in order to describe completely the preparation of a
(pure) quantum system . Given $([\psi ],\alpha ,z)$ every observable $A$
receives a univocal value $\widehat{A}([\psi ],\alpha ,z)$. Therefore all
physical quantities of the system are well defined independently from the
measuring process. Making the unknown parameter $z$ to vary in its space we
get the probabilities prescribed by quantum mechanics.We show also that
every evolution of quantum states can be extended to an evolution of the
ternes $([\psi ],\alpha ,z)$.
\end{abstract}

\section{\textbf{\ Introduction}}

Let's assign, once for all, a quantum physical system (described through the
Hilbert space $\mathbb{\mathcal{H}}$) prepared in the (pure) state $[\psi ]$
in $\mathbb{P(\mathcal{H})}$ and a measuring apparatus for the observable
corresponding to the self-adjoint operator $A$ on $\mathbb{\mathcal{H}}$.

Let's suppose to perform infinite times the measurement of $A$ on $[\psi ]$
giving to each value a different label $z$ (where $z$ varies in a set $Z$)
and to collect all the values obtained in a function $\widetilde{A}%
:Z\rightarrow \mathbb{R}$; for example the label space $Z$ could be the
Minkowski spacetime and each label $z$ be the event where/when the measured
value begins to be available.

Proceeding with the measurements, from the thickening and the thinning of
the labels in $Z$, let's suppose that a probability measure $\mu $ is
emerging informing us with the number $\mu (X)$ how probable it is to find a
label in the subset $X$ \ (with $X$ in a suitable $\sigma -$algebra $%
\mathcal{X}$ of subsets in $Z$).

The measure space $(Z,\mathcal{X,}\mu )$ so obtained allows us to compute
through the number $\mu \left[ \left( \widetilde{A}\right) ^{-1}(B)\right]
=\left( \widetilde{A}_{\ast }\mu \right) (B)$ the probability that the
measured values fall in a given borel subset $B$ of $\mathbb{R}$ and to
check if in our laboratory the probabilities furnished by the Quantum
Mechanics are verified: it is so if $\left( \widetilde{A}_{\ast }\mu \right)
(B)=\left\langle E_{B}^{A}\right\rangle _{\psi }$ for every borel subset $B$
of $\mathbb{R}$ (where $\left\{ E_{B}^{A}\right\} _{B}$ is the spectral
measure of the operator $A$).

If we introduce the cumulative monotone function $F:\mathbb{R\rightarrow }%
\left[ 0,1\right] $ defined by $F(r)=\left\langle E_{(-\infty
,r]}^{A}\right\rangle _{\psi }$ and its borel measure $\nu
_{F}(B)=\left\langle E_{B}^{A}\right\rangle _{\psi }$this condition can be
written, more synthetically, as: $\widetilde{A}_{\ast }\mu =\nu _{F}$; this
will imply also: $\left\langle A\right\rangle _{\psi }=\int_{Z}\widetilde{A}%
\cdot d\mu $, that is the attended value of QM for $A$ on $\psi $ is the
natural mean value of $\widetilde{A}$ with respect to $\mu $.

Given $F$ the condition $\widetilde{A}_{\ast }\mu =\nu _{F}$ is not enough
to reobtain a single values function $\widetilde{A}\mathbb{.}$ By experience
we know that a different laboratory will provide a different function still
verifying the probabilities of QM. Therefore we ask how many solutions the
following mathematical problem has: assigned the cumulative function $F$ and
the measure space $(Z,\mathcal{X,}\mu )$ find all borel functions $%
\widetilde{A}:Z\rightarrow \mathbb{R}$ verifying $\widetilde{A}_{\ast }\mu
=\nu _{F}$.

From a physical viewpoint to solve this problem it means to find all the
values function $\widetilde{A}$, relative to $A$ and $[\psi ]$, in all
laboratories where our scheme is appliable, starting from the function $F$
that summarizes only the attended probabilities.

If we suppose that $(Z,\mathcal{X})$ is a standard uncountable borel space
and that the probability measure $\mu $ is without atoms (that is $\mu
(\left\{ z\right\} )=0$ for every $z$ in $Z$) our problem has the following
answer: for each borel function $\alpha :Z\rightarrow ]0,1[$ (a barrier
function) such that $\alpha _{\ast }\mu =\lambda _{]0,1[}$ we get the
solution: $\widehat{A}_{\alpha }(z)=\min \left\{ r\in \mathbb{R;}\text{ }%
F(r)\geq \alpha (z)\right\} $and every solution to the problem coincides
(essentially) with one of these functions $\widehat{A}_{\alpha }$ (that we
will call observable functions of $A$ on $[\psi ]$).

We observe that , fixed the observable $A$ and the state $[\psi ]$, the
measured value $\widehat{A}_{\alpha }(z)$ depends also on the label $z$ and
on the function $\alpha $. The variable $z$ plays the role of a hidden
random variable since its variations give raise to the attended
probabilities and (in agreement with the Kochen and Specker theorem) the
measured value depends also on the function $\alpha $ that tells us how the
operator $A$ representing the measurement apparatus and the physical system
in the state $[\psi ]$ are to be connected to give the measured value.

The value $\alpha (z)$ is not hidden, in general: in fact we will see that
the function $\alpha (z)$ is worth generally: $\ \alpha (z)=F(\widehat{A}%
_{\alpha }(z))$. Therefore, fixed the space of labels and performing the
various measurements, we can build the structure $(Z,\mathcal{X,}\mu )$ with
the values function $\widehat{A}_{\alpha }$ as well as\ the basic behaviour
of the function $\alpha $.

In this context it is evident why $A$ and $[\psi ]$ are not enough to
determine a univocal measured value: you don't know the value of the
function $\alpha $ on the label $z$. That is you don't know how the
measurement apparatus and the physical system are matematically correlated.

For example in the case of the famous Schr\oe dinger's cat if $\left[ \psi %
\right] $ is the state of the system inside the box and $A$ is a test
telling if your cat is awake or not, the value $\widehat{A}_{\alpha }(z)$ is
either $1$ or $0$ according to $\alpha (z)>1-\left\langle A\right\rangle
_{\psi }$ or $\alpha (z)\leq 1-\left\langle A\right\rangle _{\psi }$.

Conversely if you know $\alpha (z)$ there is no room for uncertainty: your
measurement apparatus will show on your display the outcome:$\ \min \left\{
r\in \mathbb{R;}F(r)\geq \alpha (z)\right\} $ because the function $\alpha $
is exactly what you need to pass from $F$ to $\widehat{A}_{\alpha }$.

The insufficiency of $A$ and $[\psi ]$ induces us to propose a revision of
the measurement process: when you perform the measurement of the observable $%
A$ on a physical system prepared in the state $[\psi ]$ the outcome depends
also on the value $\alpha (z)$ taken by the ''scalar field'' $\alpha $ on
the ''hidden'' label $z$.

The condition $\alpha _{\ast }\mu =\lambda _{]0,1[}$ joined with the
hypothesis that the measure $\mu $ depends only on the state $[\psi ]$ makes
it natural to see $\alpha $ as an enrichment of the quantum state and
suggests to pass from considering the ''incomplete'' state $[\psi ]$ to
consider the ''complete'' state $([\psi ],\alpha ,z).$

Called $\mathcal{C}$ the set of all these possible ''complete states'' every
self-adjoint operator $A$ defines a function $\widehat{A}:\mathcal{%
C\rightarrow }\mathbb{R}$ by $\widehat{A}([\psi ],\alpha ,z)=\min \left\{
r\in \mathbb{R;}F(r)\geq \alpha (z)\right\} $, that is, on complete states
each observable \textbf{has a well defined value}. The theory we are
describing here is \textbf{realistic}: given the proper ingredients, the
measurement process can only produce an outcome that you can know a priori.
QM is logically compatible with a realistic description.

Moreover we show how it is possible to enrich each unitary map $\mathcal{%
U:H\rightarrow H}$ to an automorphism $\widehat{\mathcal{U}}:\mathcal{%
C\rightarrow C}$ in such a way to have a group homomorphism and moreover $%
\widehat{\mathcal{U}^{-1}A\mathcal{U}}=\widehat{A}\circ \widehat{\mathcal{U}}
$. Therefore each unitary evolution in $\mathcal{H}$ can be seen as the
apparent part of a deterministic hidden evolution in $\mathcal{C}$.

All this works for a general borel standard (uncountable) space $Z$ but if
we choose the phase space as space of labels there is a natural choice for
the measures $\mu _{\lbrack \psi ]}$ making the functions of the positions
and the functions of the momentum observable functions. Note that you need
different barriers functions for position and momentum observables.

In the end it is worth mentioning that in this context a phenomenon of
dependence of the ''state'' on the way you get the outcomes appears : there
are operators $A$ such that the operation of squaring the value $\widehat{A}%
([\psi ],\alpha ,z)$ is not given by $A^{2}$ on the ''state'' $([\psi
],\alpha ,z)$ but on a changed ''state'' $([\psi ],\beta ,z)$.

An observer reading the value $v=\widehat{A}(\left[ \psi \right] ,\alpha ,z)$
and then computing the value $v^{2}$ can continue to refer it to the
inaltered ''state'' $(\left[ \psi \right] ,\alpha ,z)$ but another observer
looking at the entire process as a measurement of the observable $A^{2}$ is
compelled to refer the outcome to the ''state'' $(\left[ \psi \right] ,\beta
,z)$ generally different from the previous one.

\section{\textbf{\ Symbols}}

In the real line we will consider the family $\mathcal{B}\left( \mathbb{R}%
\right) $ of all its borel subsets; the symbol $\chi _{B}$ will denote the
characteristic function of the subset $B$. A borel function on $\mathbb{R}$
is called here compact (bornological) if it sends bounded sets in bounded
sets: all bounded real functions and all continuous real functions are
compact.

Given a monotone, non-decreasing function $F:\mathbb{R\rightarrow \lbrack }0%
\mathbb{,}1\mathbb{]}$ if it is right-continuous with $\inf F=0$ and $\sup
F=1$ it is well defined its quasi-inverse (or quantile) function $F^{(-1)}:%
\mathbb{]}0\mathbb{,}1\mathbb{[\rightarrow }\mathbb{R}$ given by: $%
F^{(-1)}(s)=\min \left\{ r;\text{ }F(r)\geq s\right\} $. For every monotone,
non-decreasing, right-continuous function $F$ on $\mathbb{R}$ we will denote
by$\ \nu _{F}$ the borel measure characterized by $\nu _{F}]a,b]=F(b)-F(a)$.
We will denote by $\lambda _{(a,b)}=\nu _{id}$ the borel-lebesgue measure on
the interval $(a,b)$.

Given a self-adjoint operator $A$ on a Hilbert space $\mathbb{\mathcal{H}}$
with spectral family $\left\{ E_{(-\infty ,r]}^{A}\right\} _{r\in \mathbb{R}%
} $ and a unitary vector $\psi $ we will denote by $F_{\psi }^{A}$ the
monotone function defined by $F_{\psi }^{A}(r)=\left\langle E_{(-\infty
,r]}^{A}\right\rangle _{\psi }$. Note that $\nu _{F_{\psi
}^{A}}(B)=\left\langle E_{B}^{A}\right\rangle _{\psi }$ for every $B$ in $%
\mathcal{B}\left( \mathbb{R}\right) $, moreover $F_{\psi }^{A}$ is
non-decreasing, right-continuous with $\inf F_{\psi }^{A}=0$ and $\sup
F_{\psi }^{A}=1$, so it is well defined $\left( F_{\psi }^{A}\right) ^{(-1)}:%
\mathbb{]}0\mathbb{,}1\mathbb{[\rightarrow }\mathbb{R}$. Note that the
function $F_{\psi }^{A}$ is continuous in $r$ if and only if $\left\langle
E_{\left\{ r\right\} }^{A}\right\rangle _{\psi }=0$.

We will denote by $\Omega $ the family of all linear operators (essentially)
self-adjoint on $\mathbb{\mathcal{H}}$ (those associated to quantum
observables) and by $\Omega _{b}$ its subfamily of bounded operators. For
every $A$ in $\Omega $ and every borel real function $b$ the composition $%
b\circ A$ is still in $\Omega $, for every $A$ in $\Omega _{b}$ and every
compact borel real function $b$ the composition $b\circ A$ is still in $%
\Omega _{b}$.

The function $F:\Omega \times \mathbb{P(\mathcal{H})\times R\rightarrow }$ $%
\mathbb{[}0\mathbb{,}1\mathbb{]}$ given by $F(A,[\psi ],r)=F_{\psi }^{A}(r)$
resumes all the probabilistic content of a quantum system represented by the
Hilbert space $\mathbb{\mathcal{H}}$.

Given two borel spaces $(X,\mathcal{A})$ and $(Y,\mathcal{B})$ a measure $%
\mu $ on $(X,\mathcal{A})$ and a borel map $\varphi :X\rightarrow Y$ the
image measure of $\mu $ by $\varphi $ is the measure $\varphi _{\ast }\mu $
on $Y$ defined by $\left( \varphi _{\ast }\mu \right) (B)=\mu (\varphi
^{-1}(B))$. Let's remember that $\left( F_{\psi }^{A}\right) _{\ast
}^{(-1)}\lambda _{\mathbb{]}0\mathbb{,}1\mathbb{[}}=\nu _{F_{\psi }^{A}}$
(cfr. \textbf{[KS]}). In general, for any operator $A$ and any borel
function $b$ we have: $\nu _{F_{\psi }^{b\circ A}}=b_{\ast }\nu _{F_{\psi
}^{A}}$.

\section{\textbf{\ Quantum mechanics on a standard space}}

We fix a separable Hilbert complex space $\mathbb{\mathcal{H}}$ of infinite
dimension, a borel standard (uncountable) space $(Z,\mathcal{X})$ and a
probability measure without atoms $\mu _{\lbrack \psi ]}$ on $Z$ for each $%
[\psi ]$ in the projective space $\mathbb{P(\mathcal{H})}$.

What follows works for any such structure $\mathcal{(}Z\mathcal{,X,H,\mu }%
_{\cdot })$.

First of all let's guarantee that our future definitions will not be empty.

\begin{theorem}
Let $(Z,\mathcal{X})$ be a borel standard (uncountable) space

\begin{enumerate}
\item  There exists a borel equivalence between $(Z,\mathcal{X})$ and $%
(]0,1[,\mathcal{B(]}0\mathcal{,}1\mathcal{[)})$

\item  Each borel equivalence $\sigma :Z\diagdown M\rightarrow
]0,1[\diagdown N$, where $M\in \mathcal{X}$ and $N\in \mathcal{B(]}0\mathcal{%
,}1\mathcal{[)})$ with $\lambda (N)=0$, defines a probability measure $\mu
=\left( \sigma ^{-1}\right) _{\ast }\lambda _{]0,1[\diagdown N}$ without
atoms on $Z$

\item  Conversely given a probabily measure $\mu $ without atoms on $Z$
there exists a borel equivalence $\sigma :Z\rightarrow ]0,1[$ (defined out
of borel null subsets) such that $\sigma _{\ast }\mu =\lambda _{]0,1[}$.
\end{enumerate}
\end{theorem}

\begin{proof}
1. Cfr. \textbf{[KS]} Supplement to ch. 5

2. Obvious

3. Cfr \textbf{[Ry]} Thm. 9 p. 327
\end{proof}

\begin{definition}
Let's fix a self-adjoint linear operator $A$ on $\mathcal{H}$ and a quantum
state $[\psi ]$ in $\mathbb{P(\mathcal{H})}$. A \textbf{probability barrier
associated to\ }$[\psi ]$ \textbf{\ }is a borel function $\alpha
:Z\rightarrow ]0,1[$ such that $\alpha _{\ast }\mu _{\lbrack \psi ]}=\lambda
_{]0,1[}$.
\end{definition}

\begin{definition}
A \textbf{complete (pure) state} on $\mathcal{(}Z\mathcal{,X,H,\mu }_{\cdot
})$ is a triple $(\left[ \psi \right] ,\alpha ,z)$ of a pure state $\left[
\psi \right] $ in $\mathbb{P(\mathcal{H})}$, a barrier $\alpha _{\left[ \psi %
\right] }$ associated to $\left[ \psi \right] $ and a point $z$ in $Z$.
\end{definition}

\begin{notation}
Let's denote by $\mathcal{C}$ the set of all complete states on $\mathcal{(}Z%
\mathcal{,H,\mu }_{\cdot })$.
\end{notation}

\begin{definition}
The \textbf{values function} $v:\Omega \times \mathcal{C\rightarrow }\mathbb{%
R}$ is defined by:
\begin{equation*}
v(A;\left[ \psi \right] ,\alpha _{\lbrack \psi ]},z)=\min \left\{ r\in
\mathbb{R:}\text{ }\left\langle E_{(-\infty ,r]}^{A}\right\rangle _{\psi
}\geqq \alpha _{\lbrack \psi ]}(z)\right\} =\left( F_{\psi }^{A}\right)
^{(-1)}\left( \alpha _{\lbrack \psi ]}(z)\right)
\end{equation*}
\end{definition}

\begin{notation}
Fixed a self-adjoint linear operator $A$ on $\mathcal{H}$ its \textbf{values
function} is the function $\widehat{A}:\mathcal{C\rightarrow }\mathbb{R}$
defined by: $\widehat{A}(\left[ \psi \right] ,\alpha _{\lbrack \psi
]},z)=v(A;\left[ \psi \right] ,\alpha _{\lbrack \psi ]},z)$. Fixed also a
pure state $\left[ \psi \right] $ and a barrier $\alpha _{\lbrack \psi ]}$%
associated to $\left[ \psi \right] $ we will denote by $\widehat{A}_{\alpha %
\left[ \psi \right] }:Z\rightarrow \mathbb{R}$ the function $\widehat{A}$
with $\left[ \psi \right] $ and $\alpha _{\lbrack \psi ]}$ fixed.
\end{notation}

\begin{theorem}
\begin{enumerate}
\item  $\widehat{A}_{\alpha \left[ \psi \right] }=\left( F_{\psi
}^{A}\right) ^{(-1)}\circ \alpha _{\lbrack \psi ]}$ is a borel function

\item  $\widehat{A}_{\alpha \left[ \psi \right] \ast }\mu _{\left[ \psi %
\right] }=\nu _{F_{\psi }^{A}}$.
\end{enumerate}
\end{theorem}

\begin{proof}
\begin{enumerate}
\item  A quasi-inverse function is monotonic.

\item  $\widehat{A}_{\alpha \left[ \psi \right] \ast }\mu _{\left[ \psi %
\right] }=\left( \left( F_{\psi }^{A}\right) ^{(-1)}\circ \alpha _{\lbrack
\psi ]}\right) _{\ast }\mu _{\left[ \psi \right] }=\left( F_{\psi
}^{A}\right) _{\ast }^{(-1)}\lambda _{]0,1[}=\nu _{F_{\psi }^{A}}$.
\end{enumerate}
\end{proof}

\begin{remark}
The property $\widehat{A}_{\alpha \left[ \psi \right] \ast }\mu _{\left[
\psi \right] }=\nu _{F_{\psi }^{A}}$ means that for each borel subset $B$ in
$\mathbb{R}$ the probability $\mu _{\left[ \psi \right] }\left( \left(
\widehat{A}_{\alpha \left[ \psi \right] }\right) ^{-1}(B)\right) $ that the
value of the function $\widehat{A}_{\alpha \left[ \psi \right] }$ falls in $%
B $ is the probability $\nu _{F_{\psi }^{A}}(B)=\left\langle
E_{B}^{A}\right\rangle _{\psi }$ assigned by the Quantum Mechanics for $A$, $%
\left[ \psi \right] $ and $B$.
\end{remark}

\begin{exercise}
\begin{itemize}
\item  For each projector $E$ on $\mathcal{H}$ we have: $\widehat{\left(
E\right) }_{\alpha \left[ \psi \right] }(z)=\chi _{L\text{ }_{\left[ \psi %
\right] }}(z)$ with $L_{\left[ \psi \right] }=\left\{ z\in Z;1-\left\langle
E\right\rangle _{\psi }<\alpha _{\left[ \psi \right] }(z)<1\right\} $

\item  Let $\left\{ E_{0},...,E_{m,}...,E_{\infty }\right\} $ be a sequence
of pairwise orthogonal projectors in $\mathcal{H}$ such that $%
I=\sum_{k}E_{k} $\ \ and let $\lambda _{0},...,\lambda _{m},...,\lambda
_{\infty }$ be a sequence of real numbers with $\lambda _{\infty }=0$, $%
\left\{ \lambda _{m}\right\} _{m\geqslant 0}$ strictly increasing to $0$ and
defining the bounded operator $A=\sum_{k}\lambda _{k}\cdot E_{k}$, its
function $\widehat{A}_{\alpha \left[ \psi \right] }$ is the $\sigma $-simple
function $\widehat{A}_{\alpha \left[ \psi \right] }=\sum_{k}\lambda
_{k}\cdot \chi _{L_{k\text{ }\left[ \psi \right] }}$where:
\begin{eqnarray*}
L_{k\left[ \psi \right] } &=&\left\{ z\in Z;\sum_{j=0,...,k-1}\left\langle
E_{j}\right\rangle _{\psi }<\alpha _{\left[ \psi \right] }(z)\leq
\sum_{j=0,...,k}\left\langle E_{j}\right\rangle _{\psi }\right\} \\
L_{\infty \left[ \psi \right] } &=&\left\{ z\in
Z;\sum_{j=0,...,k}\left\langle E_{j}\right\rangle _{\psi }<\alpha _{\left[
\psi \right] }(z)\text{ for each }k<\infty \right\}
\end{eqnarray*}
Note that\ the family $\left\{ L_{k\left[ \psi \right] }\right\} $ makes a
partition of $Z$.
\end{itemize}
\end{exercise}

\begin{theorem}
When $A$ is a bounded self-adjoint operator, a function $f:Z\rightarrow
\mathbb{R}$ borel and essentially bounded for $\mu _{\left[ \psi \right] }$%
verifies $f_{\ast }\mu _{\left[ \psi \right] }=\nu _{F_{\psi }^{A}}$ if and
only if for each borel compact function $b$ and each unitary vector $\psi $,
we have:
\begin{equation*}
\int_{Z}\left( b\circ f\right) (z)\cdot d\mu _{\left[ \psi \right]
}=\left\langle \psi ,\left( b\circ A\right) \psi \right\rangle
\end{equation*}
\end{theorem}

\begin{proof}
$\left[ \Longrightarrow \right] $ Since $f$ is essentially bounded for $\mu
_{\left[ \psi \right] }$ and $b$ is compact the function $b\circ f$ is
essentially bounded for $\mu _{\left[ \psi \right] }$ and absolutely
integrable, moreover: $\int_{Z}\left( b\circ f\right) (z)\cdot d\mu _{\left[
\psi \right] }=\int_{\mathbb{R}}b\cdot d(f_{\ast }\mu _{\left[ \psi \right]
})=\int_{\mathbb{R}}b\cdot d(\nu _{F_{\psi }^{A}})=\left\langle b\circ
A\right\rangle _{\psi }$

$\left[ \Longleftarrow \right] $For each borel subset $B$ in $\mathbb{R}$ we
have $\left( f_{\ast }\mu _{\left[ \psi \right] }\right) (B)=\int_{\mathbb{R}%
}\chi _{B}\cdot d(f_{\ast }\mu _{\left[ \psi \right] })=\int_{Z}\left( \chi
_{B}\circ f\right) \cdot d\mu _{\left[ \psi \right] }=\left\langle \chi
_{B}\circ A\right\rangle _{\psi }=\nu _{F_{\psi }^{A}}(B)$
\end{proof}

\begin{notation}
Introducing the symbol $\left\langle \left\langle b\circ f\right\rangle
\right\rangle _{\psi }=\int_{Z}b\circ f\cdot d\mu _{\left[ \psi \right] }$
we can write $\left\langle \left\langle b\circ f\right\rangle \right\rangle
_{\psi }=\left\langle b\circ A\right\rangle _{\psi }$. In particular $%
\left\langle \left\langle f^{k}\right\rangle \right\rangle _{\psi
}=\left\langle A^{k}\right\rangle _{\psi }$ for every $k\geq 0$.
\end{notation}

\begin{remark}
The next property even though it appears quite often in synthetic data
generation it deserves to be full proved in our context.
\end{remark}

\begin{lemma}
Let $F:\mathbb{R\rightarrow \lbrack }0\mathbb{,}1\mathbb{]}$ be a monotone,
non-decreasing and right continuous function with $\inf F(\mathbb{R})=0$ and
$\sup F(\mathbb{R})=1$. Let $f:]0,1[\rightarrow \mathbb{R}$ a borel function
such that $f_{\ast }\lambda _{]0,1[}=\nu _{F}$, there exist two borel null
subsets $M$ and $N$ in $]0,1[$ and a borel function $\alpha :]0,1[\setminus
M\rightarrow ]0,1[\setminus N$ such that $\alpha _{\ast }\lambda
_{]0,1[\setminus M}=\lambda _{]]0,1[\setminus N}$ and $f\mid
_{]0,1[\setminus M}=F^{(-1)}\circ \alpha $.
\end{lemma}

\begin{proof}
The function $F$ can have a countable set $\left\{ x_{k}\right\} _{k\geq 1}$
of discontinuity points; let's write $a_{k}=F^{-}(x_{k})<F(x_{k})=b_{k}$.
Let's denote by $][a_{k},b_{k}[$ the interval $[a_{k},b_{k}[\setminus
(\left\{ a_{k}\right\} \cap F(\mathbb{R}))$.

Moreover the function $F$ can be constant on a countable family of disjoint
proper maximal intervals $\left\{ J_{n}\right\} _{n\geq 1}$. Let's write $%
F(J_{n})=\left\{ c_{n}\right\} $. We have $F^{-1}(F(r))=J_{n}$ if $r\in
J_{n} $ and instead $F^{-1}(F(r))=\left\{ r\right\} $ if $r\notin \bigcup
J_{n}$. If $a_{k}\in F(\mathbb{R})$ there exists $J_{n}$ such that $\left\{
a_{k}\right\} =F(J_{n})=\left\{ c_{n}\right\} $.

The restricted function $F\mid :\left( \mathbb{R\setminus }\bigcup
J_{n}\right) \rightarrow \left( ]0,1[\setminus \left[ \bigcup
][a_{k},b_{k}[\cup \left\{ c_{n}\right\} \right] \right) $ is a bijective
borel equivalence with $\left( F\mid \right) ^{-1}=F^{(-1)}\mid $(observe
that $]0,1[\setminus \bigcup ][a_{k},b_{k}[\subset F(\mathbb{R})$). \ Taken $%
M_{0}=f^{-1}(\bigcup J_{n}\cup \left\{ x_{k}\right\} )$ and $N_{0}=\bigcup
][a_{k},b_{k}]\cup \left\{ c_{n}\right\} $ we have $\lambda (M_{0})=\lambda
(N_{0})=0$ and the function $\alpha _{0}=F\circ f:]0,1[\setminus
M_{0}\rightarrow ]0,1[\setminus N_{0}$ is a borel function with $%
F^{(-1)}\circ \alpha _{0}=f$ and $\alpha _{0\ast }\lambda =\lambda $.

Since $\lambda \left[ f^{-1}(\left\{ x_{k}\right\} )\right]
=b_{k}-a_{k}=\lambda ][a_{k},b_{k}[$ for each $k\geq 1$ there exists a borel
null subset $M_{k}$ of $f^{-1}(\left\{ x_{k}\right\} )$ and a borel null
subset $N_{k}$ of $][a_{k},b_{k}[$ and a borel and measure equivalence $%
\alpha _{k}:f^{-1}(\left\{ x_{k}\right\} )\setminus M_{k}\rightarrow
][a_{k},b_{k}[\setminus N_{k}$.

Let's take $M=\bigcup_{n\geq 0}M_{n}$ and $N=\bigcup_{n\geq 0}N_{n}$, the
map $\alpha :]0,1[\setminus M\rightarrow ]0,1[\setminus N$ defined by $%
\alpha _{0}=F\circ f$ on $]0,1[\setminus f^{-1}(\bigcup J_{n}\cup \left\{
x_{k}\right\} )$ and $a_{k}$ on $f^{-1}(\left\{ x_{k}\right\} )\setminus
M_{k}$ verifies the thesis of the lemma.
\end{proof}

\begin{theorem}
A borel function $f_{_{\left[ \psi \right] }}:Z\rightarrow \mathbb{R}$
verifies $f_{_{\left[ \psi \right] }\ast }\mu _{\left[ \psi \right] }=\nu
_{F_{\psi }^{A}}$ if and only if there exists a barrier $\alpha _{\lbrack
\psi ]}$ associated to $\left[ \psi \right] $ such that $f_{_{\left[ \psi %
\right] }}=\left( F_{\psi }^{A}\right) ^{(-1)}\circ \alpha _{\lbrack \psi ]}$
(out of a borel $\mu _{\left[ \psi \right] }-$null subset).
\end{theorem}

\begin{proof}
$\left[ \Longrightarrow \right] $ We already proved that $\left[ \left(
F_{\psi }^{A}\right) ^{(-1)}\circ \alpha _{\lbrack \psi ]}\right] _{\ast
}\mu _{\left[ \psi \right] }=\nu _{F_{\psi }^{A}}$.

$\left[ \Longleftarrow \right] $Since $Z$ is an uncountable borel standard
space we can find a borel $\mu _{\left[ \psi \right] }-$null subspace $M$ in
$Z$, a borel null subspace $N$ in $]0,1[$ and a borel equivalence $\sigma
:Z\diagdown M\rightarrow ]0,1[\diagdown N$ with $\sigma _{\ast }\mu _{\left[
\psi \right] }=\lambda _{]0,1[}$. Therefore the function $f\circ \sigma
^{-1}:]0,1[\diagdown N\rightarrow \mathbb{R}$ verifies: $\left( f\circ
\sigma ^{-1}\right) _{\ast }\lambda _{]0,1[}=\nu _{F_{\psi }^{A}}$.

For the previous Lemma there exists a borel function $\beta
:]0,1[\rightarrow ]0,1[$ with $\beta _{\ast }\lambda _{]0,1[}=\lambda
_{]0,1[}$ and $f\circ \sigma ^{-1}=\left( F_{\psi }^{A}\right) ^{(-1)}\circ
\beta $ out of a borel null subset, therefore $f=\left( F_{\psi }^{A}\right)
^{(-1)}\circ \alpha _{\left[ \psi \right] }^{A}$ out of a borel null subset
where $\alpha _{\left[ \psi \right] }^{A}=\beta \circ \sigma $ is a barrier
associated to $\left[ \psi \right] $.
\end{proof}

\begin{remark}
From the definition of quasi-inverse function for $f_{_{\left[ \psi \right]
}}=\left( F_{\psi }^{A}\right) ^{(-1)}\circ \alpha _{\lbrack \psi ]}$ we
have: \ \ $F_{\psi }^{A}(f_{_{\left[ \psi \right] }}(z)^{-})\leq \alpha
_{\lbrack \psi ]}(z)\leq F_{\psi }^{A}(f_{_{\left[ \psi \right] }}(z))$.
Therefore if $f_{_{\left[ \psi \right] }}(z)$ is not a discontinuity point
of $F_{\psi }^{A}$ we have: $\alpha _{\lbrack \psi ]}(z)=F_{\psi }^{A}(f_{_{%
\left[ \psi \right] }}(z))$.
\end{remark}

\section{\textbf{\ Representations of bounded operators by functions}}

\begin{definition}
A \textbf{complex of barriers} is a family $\alpha =$ $\left\{ \alpha
_{\lbrack \psi ]}^{A}\right\} _{A\text{ }\in \Omega \text{,}\left[ \psi %
\right] \in \mathbb{P(\mathcal{H})}}$ where each $\alpha _{\lbrack \psi
]}^{A}$ is a barrier for $\left[ \psi \right] $.
\end{definition}

\begin{notation}
Fixed a complex of barriers $\alpha $ for each self-adjoint operator $A$ in $%
\Omega $ it is defined the function $\widehat{A}_{\alpha }:\mathbb{P(%
\mathcal{H})\times }Z\rightarrow \mathbb{R}$ by $\widehat{A}_{\alpha }([\psi
],z)=v(A;[\psi ],\alpha _{\lbrack \psi ]}^{A},z)=\left( \left( F_{\psi
}^{A}\right) ^{(-1)}\circ \alpha _{\lbrack \psi ]}^{A}\right) (z)$ . Two
such functions $\widehat{H}_{\alpha }$ and $\widehat{K}_{\beta }$will be
considered \textbf{equivalent} if, for each $\left[ \psi \right] $ in $%
\mathbb{P(\mathcal{H})}$ the functions $\widehat{H}_{\alpha \left[ \psi %
\right] }$ and $\widehat{K}_{\beta \left[ \psi \right] }$ are equal out of a
borel subset $\mu _{\left[ \psi \right] }$-null.
\end{notation}

\begin{notation}
Given two self-adjoint linear operators $A_{1}$ and $A_{2}$ and a complex of
barriers $\alpha $ if $\widehat{A}_{1\alpha }\equiv \widehat{A}_{2\alpha }$
then $A_{1}=A_{2}$.
\end{notation}

\begin{proof}
For each pure state $\left[ \psi \right] $ from $\widehat{A}_{1\alpha \left[
\psi \right] }\equiv \widehat{A}_{2\alpha \left[ \psi \right] }$ it follows $%
\nu _{F_{\psi }^{A_{1}}}=\nu _{F_{\psi }^{A_{2}}}$ and therefore $F_{\psi
}^{A_{1}}=F_{\psi }^{A_{2}}$ since both are right-continuous functions with $%
\inf =0$ and $\sup =1$.

Then $\left\langle E_{(-\infty ,r]}^{A_{1}}\right\rangle _{\psi
}=\left\langle E_{(-\infty ,r]}^{A_{2}}\right\rangle _{\psi }$ for each $r$
in $\mathbb{R}$ and each unitary vector $\psi $. That is $A_{1}=A_{2}$.
\end{proof}

\begin{lemma}
For each $A$ in $\Omega _{b}$ and each $\left[ \psi \right] $ in $\mathbb{P(%
\mathcal{H})}$: $\mu _{\left[ \psi \right] }\left[ (\widehat{A}_{\alpha %
\left[ \psi \right] })^{-1}]a,b]\right] =0$ if and only if $(\widehat{A}%
_{\alpha \left[ \psi \right] })^{-1}]a,b]=\varnothing $.
\end{lemma}

\begin{proof}
Since $\widehat{A}_{\alpha \left[ \psi \right] \ast }\mu _{\left[ \psi %
\right] }=\nu _{F_{\psi }^{A}}$ we have $\mu _{\left[ \psi \right] }\left[ (%
\widehat{A}_{\alpha \left[ \psi \right] })^{-1}]a,b]\right] =0$ if and only
if $F_{\psi }^{A}(a)=F_{\psi }^{A}(b)$ that is equivalent to $\left[ \left(
F_{\psi }^{A}\right) ^{(-1)}\right] ^{-1}]a,b]=\varnothing $ that implies $(%
\widehat{A}_{\alpha \left[ \psi \right] })]a,b]=\varnothing $.
\end{proof}

\begin{theorem}
For each $A$ in $\Omega _{b}$ we have $\overline{\widehat{A}_{\alpha }(%
\mathbb{P(\mathcal{H})\times }Z)}=spec(A)$.
\end{theorem}

\begin{proof}
For a number $r$ in $\mathbb{R}$ it holds $r\notin spec(A)$ if and only if
there exists $\sigma >0$ with $E_{]r-\sigma ,r+\sigma ]}^{A}=0$ (cfr.
\textbf{[W]} Thm. 7.22) that is equivalent to $\mu _{\left[ \psi \right] }%
\left[ \widehat{A}_{\alpha \left[ \psi \right] }^{-1}]r-\sigma ,r+\sigma ]%
\right] =0$ and to $\widehat{A}_{\alpha \left[ \psi \right] }^{-1}]r-\sigma
,r+\sigma ]=\varnothing $ for each $\left[ \psi \right] $ in $\mathbb{P(%
\mathcal{H})}$ that is $r\notin \overline{\widehat{A}_{\alpha }(\mathbb{P(%
\mathcal{H})\times }Z)}$.
\end{proof}

\begin{definition}
A\textbf{\ representation of }$\Omega _{b}$\textbf{\ with functions of }$%
\mathcal{(}Z\mathcal{,X,\mu }_{\cdot })$ is a map $\widetilde{(\cdot )}%
:\Omega _{b}\rightarrow \mathbb{R}^{\mathbb{P(\mathcal{H})\times }Z}$ such
that for each $A$ in $\Omega _{b}$ and for each $\left[ \psi \right] $ in $%
\mathbb{P(\mathcal{H})}$:

\begin{enumerate}
\item  the function $\widetilde{A}_{\left[ \psi \right] }:Z\rightarrow
\mathbb{R}$ is borel and essentially bounded for $\mu _{\left[ \psi \right]
} $

\item  $\widetilde{A}_{\left[ \psi \right] \ast }\mu _{\left[ \psi \right]
}=\nu _{F_{\psi }^{A}}$
\end{enumerate}
\end{definition}

\begin{theorem}
Given a representation $\widetilde{(\cdot )}:\Omega _{b}\rightarrow \mathbb{R%
}^{\mathbb{P(\mathcal{H})\times }Z}$\textbf{\ }of\textbf{\ }$\Omega _{b}$%
\textbf{\ }on\textbf{\ }$\mathcal{(}Z\mathcal{,X,\mu }_{\cdot })$ there
exists a complex of barriers $\alpha $ such that $\widetilde{A}\equiv
\widehat{A}_{\alpha }$.
\end{theorem}

\begin{proof}
Given a representation $\widetilde{(\cdot )}:\Omega _{b}\rightarrow \mathbb{R%
}^{\mathbb{P(\mathcal{H})\times }Z}$ for each $A$ and $\left[ \psi \right] $
we have: $\widetilde{A}_{\left[ \psi \right] \ast }\mu _{\left[ \psi \right]
}=\nu _{F_{\psi }^{A}}$. This implies there exists a barrier $\alpha _{\left[
\psi \right] }^{A}$ associated to $\left[ \psi \right] $such that $%
\widetilde{A}_{\left[ \psi \right] }\equiv \left( F_{\psi }^{A}\right)
^{(-1)}\circ \alpha _{\lbrack \psi ]}^{A}=\widehat{A}_{\alpha \left[ \psi %
\right] }$.
\end{proof}

\begin{theorem}
A map $\widetilde{(\cdot )}:\Omega _{b}\rightarrow \mathbb{R}^{\mathbb{P(%
\mathcal{H})\times }Z}$ with each function $\widetilde{A}_{\left[ \psi %
\right] }:Z\rightarrow \mathbb{R}$ borel and $\mu _{\left[ \psi \right] }-$%
essentially bounded for each $\left[ \psi \right] $ in $\mathbb{P(\mathcal{H}%
)}$, is a representation of $\Omega _{b}$ if and only if for each bounded
self-adjoint operator $A$ on $\mathcal{H}$, for each $\left[ \psi \right] $
in $\mathbb{P(\mathcal{H})}$ and for each borel compact function $b$ we
have:
\begin{equation*}
\int_{Z}\left( b\circ \widetilde{A}_{\left[ \psi \right] }\right) (z)\cdot
d\mu _{\left[ \psi \right] }=\left\langle \psi ,\left( b\circ A\right) \psi
\right\rangle \text{ (when }\left\| \psi \right\| =1\text{)}
\end{equation*}
\end{theorem}

\begin{remark}
Given a representation $\widetilde{(\cdot )}$ of $\Omega _{b}$ for every
self-adjoint bounded operator $A$ and every borel compact function $b:%
\mathbb{R\rightarrow R}$ the attended value $\left\langle b\circ
A\right\rangle _{\psi }$ is equal to the mean value $\int_{Z}\left( b\circ
A_{\left[ \psi \right] }\right) (z)\cdot d\mu _{\left[ \psi \right] }$ and
also to the mean value $\int_{Z}\widetilde{(b\circ A)}_{\left[ \psi \right]
}(z)\cdot d\mu _{\left[ \psi \right] }$suggesting that it could be $%
\widetilde{(b\circ A)}_{\left[ \psi \right] }=b\circ \widetilde{A}_{\left[
\psi \right] }$. This is true, for example, if $b:\mathbb{R\rightarrow R}$
is a strictly increasing homeomorphism: in fact in this case we have (cfr.
\textbf{[W]}, 7.3): $F_{\psi }^{b\circ A}=F_{\psi }^{A}\circ b^{-1}$, $%
\left( F_{\psi }^{b\circ A}\right) ^{(-1)}=b\circ \left( F_{\psi
}^{A}\right) ^{(-1)}$ and then $\widehat{\left( b\circ A\right) }_{\alpha
}=b\circ \widehat{A}_{\alpha }$ for every self-adjoint operator $A$ and
every complex of barriers $\alpha $. However the next theorem tells us that
this is not universally true whatever representation you consider.
\end{remark}

\begin{theorem}
\begin{enumerate}
\item  It does not exist a complex of barriers $\alpha $ such that $b\circ
\widehat{A}_{\alpha }=$ $\widehat{\left( b\circ A\right) }_{\alpha }$ for
every self-adjoint bounded operator $A$ and every compact borel function $b$.

\item  For every self-adjoint bounded operator $A$ and every compact borel
function $b$ there exists a suitable complex $\beta $ of barriers such that $%
b\circ \widehat{A}_{\alpha }\equiv \widehat{\left( b\circ A\right) }_{\beta
} $
\end{enumerate}
\end{theorem}

\begin{proof}
\begin{enumerate}
\item  For every projector $E$ the function $\widehat{E}_{\alpha }$ is a
characteristic function. Fixed $\left[ \psi _{0}\right] $ in $\mathbb{P(%
\mathcal{H})}$ with $\left\| \psi _{0}\right\| =1$ and $z_{0}$ in $Z$ it is
well defined the function $G:\left\{ \mathbb{\psi ;}\text{ }\psi \text{ is
unitary}\right\} \mathbb{\rightarrow }\left\{ 0\mathbb{,}1\right\} $ by $%
G(\psi )=\widehat{\left( E_{\psi }\right) }_{\alpha }(\left[ \psi _{0}\right]
,z_{0})=\chi _{\left\{ 1-\alpha _{\lbrack \psi _{0}]}(z_{0})<\left|
\left\langle \psi ,\psi _{0}\right\rangle \right| ^{2}\right\} }(\psi )$.
Let's make the extra hypothesis that such a complex of barriers $\alpha $
would esist, it will imply that for the function $G$ we have: $\sum_{_{k\geq
1}}G(\psi _{k})=1$ for every orthonormal basis $\left\{ \psi _{k}\right\}
_{k\geq 1}$. We can choose a suitable sequence $\left\{ c_{k}\right\}
_{k\geq 1}$ of distinct real numbers in such a way that the self-adjoint
operator $A=\sum_{_{k\geq 1}}c_{k}\cdot E_{\psi _{k}}$ is bounded. We have
for the extra hypothesis: $1=\widehat{I}_{\alpha }=\widehat{\left(
\sum_{_{k\geq 1}}E_{\psi _{k}}\right) }_{\alpha }=\widehat{\left[ \left(
\sum_{_{k\geq 1}}\chi _{c_{k}}\right) \circ A\right] }_{\alpha }=\left(
\sum_{_{k\geq 1}}\chi _{c_{k}}\right) \circ \widehat{A}_{\alpha }$ therefore
$\sum_{_{k\geq 1}}\widehat{\left( E_{\psi _{k}}\right) }_{\alpha
}=\sum_{_{k\geq 1}}\widehat{\left( \chi _{c_{k}}\circ A\right) }_{\alpha
}=\sum_{_{k\geq 1}}\left( \chi _{c_{k}}\circ \widehat{A}_{\alpha }\right)
=\left( \sum_{_{k\geq 1}}\chi _{c_{k}}\right) \circ \widehat{A}_{\alpha }=1$%
. Then $G$ is a Gleason frame function of weight 1(cfr. \textbf{[G]}) and
there exists a bounded self-adjoint operator $D$ such that $G(\psi
)=\left\langle \psi ,D\psi \right\rangle $ for every unitary $\psi $. This
implies that $G$ is continuous and constantly $0$ or constantly $1$,
contradicting, in both cases, $\sum_{_{k\geq 1}}G(\psi _{k})=1$ for any
orthonormal basis $\left\{ \psi _{k}\right\} _{k\geq 1}$.

\item  For the function $\left( b\circ \widehat{A}_{\alpha }\right) _{\left[
\psi \right] }$we have $\left( b\circ \widehat{A}_{\alpha }\right) _{\left[
\psi \right] \ast }\mu _{\left[ \psi \right] }=b_{\ast }\nu _{F_{\psi
}^{A}}=\nu _{F_{\psi }^{b\circ A}}$. In fact: $\nu _{F_{\psi }^{b\circ
A}}(B)=\left\langle E_{B}^{b\circ A}\right\rangle _{\psi }=\left\langle \chi
_{B}^{{}}(b\circ A)\right\rangle _{\psi }=\left\langle \chi
_{b^{-1}(B)}^{{}}(A)\right\rangle _{\psi }=\nu _{F_{\psi }^{A}}(b^{-1}(B))$
(cfr. \textbf{[W] }Ch. 7.3).
\end{enumerate}
\end{proof}

\begin{remark}
. When $b\circ \widehat{A}_{\alpha }\equiv \widehat{\left( b\circ A\right) }%
_{\beta }\neq \widehat{\left( b\circ A\right) }_{\alpha }$ the composition
with the function $b$ compels us to consider a new complex of barriers $%
\beta $, different from the previous $\alpha $ used to define $\widehat{A}%
_{\alpha }$. An observer reading the value $v=\widehat{A}(\left[ \psi \right]
,\alpha _{\left[ \psi \right] }^{A},z)$ and then computing the value $b(v)$
can continue to refere it to the inaltered ''state'' $(\left[ \psi \right]
,\alpha _{\left[ \psi \right] }^{A},z)$ but another observer looking at the
entire process as a measurement of the observable $b\circ A$ seems compelled
to refere the outcome to the ''state'' $(\left[ \psi \right] ,\beta _{\left[
\psi \right] }^{A},z)$ generally different from the previous one. A possible
expanation could be that the ''state'' of a physical system can depend on
the way the outcomes emerge (cfr. \textbf{[Ro]), }however other
erxplanations are possible.
\end{remark}

\begin{example}
Let's take $Z=]0,1[$ and $\mu _{\lbrack \psi ]}=\lambda $ for every $\left[
\psi \right] $ in $\mathbb{P(\mathcal{H})}$. Let's decompose the space\ $%
\mathcal{H}$ as $\mathcal{H}\mathbb{\mathcal{=}}\mathcal{I}\mathbb{\mathcal{%
\oplus }}\mathcal{J}\mathbb{\mathcal{\oplus }}\mathcal{K}$ with $\mathcal{I}$%
, $\mathcal{J}$ and $\mathcal{K}$ orthogonal two by two and let's denote by $%
E$, $F$ and $G$ the orthogonal projections, respectively, on $\mathcal{I}$, $%
\mathcal{J}$ and $\mathcal{K}$ . Let's take a unitary vector $\psi $ such
that $\left\langle E\right\rangle _{\psi }=1/8$, $\left\langle
F\right\rangle _{\psi }=1/4$ and $\left\langle G\right\rangle _{\psi }=5/8$
and let's consider $A=E-G$ (with $A^{2}=E+G$). It is not difficult to
compute $\left[ \left( F_{\psi }^{A}\right) ^{(-1)}\right] ^{2}(s)=\chi
_{]0,5/8[\cup ]7/8,1[}(s)$ and $\left[ \left( F_{\psi }^{A^{2}}\right)
^{(-1)}\right] (s)=\chi _{]1/4,1[}(s)$. This means $\left( \widehat{A^{2}}%
\right) _{\beta }=\chi _{]1/4,1[}(\beta (s))$ and $\left( \widehat{A}%
_{\alpha }\right) ^{2}=\chi _{]0,5/8]\cup ]7/8,1[}(\alpha (s))$ so $\left(
\widehat{A}_{\alpha }\right) ^{2}\equiv \left( \widehat{A^{2}}\right)
_{\beta }$if we take $\beta (s)\equiv \sigma (\alpha (s))$ where $\sigma
(u)=(u+3/8)$ $MOD$ $1$.
\end{example}

\section{\textbf{\ Observable functions}}

\begin{definition}
A function $f:\mathbb{P(\mathcal{H})\times }Z\rightarrow \mathbb{R}$ will be
called an \textbf{observable function} if each $f_{\left[ \psi \right]
}:Z\rightarrow \mathbb{R}$ is a borel function and there exists a
self-adjoint operator $A$ on $\mathbb{\mathcal{H}}$ such that $f_{\left[
\psi \right] \ast }\mu _{\left[ \psi \right] }=\nu _{F_{\psi }^{A}}$ for
every $\left[ \psi \right] $ in $\mathbb{P(\mathcal{H})}$.
\end{definition}

\begin{remark}
The observable functions associated to an operator $A$ emulate on each $%
\left[ \psi \right] $ the values taken by the concrete apparatuses measuring
the observable $A$ on $\left[ \psi \right] $.
\end{remark}

\begin{remark}
For every self-adjoint operator $A$ and every complex of barriers $\alpha $
the function $\widehat{A}_{\alpha }$ is an observable function. Conversely
for every observable function $f$ there exist a self-adjoint operator $A$
and a complex of barriers $\alpha $ such that $f_{\left[ \psi \right]
}\equiv \widehat{A}_{\alpha \left[ \psi \right] }$ for every $\left[ \psi %
\right] $ in $\mathbb{P(\mathcal{H})}$.
\end{remark}

\begin{notation}
Let's denote by $\mathcal{O}$ the set of all the observable functions and by
$Op:\mathcal{O\rightarrow }\Omega $ the well defined map sending an
observable function to its associated operator; we will denote by $\mathcal{O%
}_{b}$ the subset of $\mathcal{O}$ of all observables whose operator is
bounded. Fixed a complex of barriers $\alpha $ we will denote by $\mathcal{O}%
_{\alpha }$ the family of all the functions $\widehat{A}_{\alpha }$ with $A$
a self-adjoint operator and by $\mathcal{O}_{\alpha b}$ the family of
functions $\widehat{A}_{\alpha }$ with $A$ a bounded self-adjoint operator.
We have $\mathcal{O=}\bigcup_{\alpha }\mathcal{O}_{\alpha }$ and $\mathcal{O}%
_{b}\mathcal{=}\bigcup_{\alpha }\mathcal{O}_{\alpha b}$.
\end{notation}

\begin{theorem}
For each observable function $f$ and each borel function $b$ the composition
$b\circ f$ is an observable function with $Op(b\circ f)=b\circ Op(f)$.
\end{theorem}

\begin{proof}
If $f$ is in $\mathcal{O}$ and $b$ is borel then each $\left( b\circ
f\right) _{\left[ \psi \right] }=b\circ f_{\left[ \psi \right] }$ is borel
and: $(b\circ f)_{\left[ \psi \right] \ast }\mu _{\left[ \psi \right]
}=b_{\ast }\nu _{F_{\psi }^{Op(f)}}=\nu _{F_{\psi }^{b\circ Op(f)}}$.
\end{proof}

\begin{theorem}
Let's consider the Hilbert space $\mathcal{H=}\bigoplus_{s=-\sigma
,...,\sigma }L^{2}(\mathcal{\mathbb{R}}_{s}^{3},\lambda )$ of a particle of
spin $\sigma $, $\ $the standard borel space $Z\mathcal{=}\bigcup_{s}%
\mathcal{\mathbb{R}}_{s}^{6}$ and the probability measures: $\mu _{\lbrack
\psi ]}=\sum_{\left\| \psi _{s}\right\| \neq 0}\frac{1}{\left\| \psi
_{s}\right\| ^{2}}\cdot \left| \psi _{s}\right| ^{2}\otimes \left| \mathcal{F%
}\psi _{s}\right| ^{2}\cdot \lambda _{\mathcal{\mathbb{R}}_{s}^{6}}$ (when $%
\left\| \psi \right\| =1$).

\begin{enumerate}
\item  For every borel function $f$ the function $f(p_{1},p_{2},p_{3})$ is
an observable function with operator $f(P_{1},P_{2},P_{3})$

\item  For every borel function $g$ the function $g(q_{1},q_{2},q_{3})$ is
an observable function with operator $g(Q_{1},Q_{2},Q_{3})$

\item  The function $s$ is an observable function with operator $S$ (defined
by $S((\psi _{s})_{s})=\left( s\cdot \psi _{s}\right) _{s}$)
\end{enumerate}
\end{theorem}

\begin{proof}
\begin{enumerate}
\item  Applying the Thm. 7.17 of \textbf{[W]} to the operator $%
g(Q_{1},Q_{2},Q_{3})$ it is not difficult to prove that $E_{(-\infty
,r]}^{g(Q)}\psi =\chi _{\left\{ g(q)\leq r\right\} }\cdot \psi $ and $%
F_{\psi }^{g(Q)}(r)=\sum_{\left\| \psi _{s}\right\| \neq 0}\iint_{\mathcal{%
\mathbb{R}}^{6}}\chi _{\left\{ g(q)\leq r\right\} }\cdot \left| \psi
_{s}\right| ^{2}\cdot dqdp$. Therefore $\left( g(q)_{\ast }\mu _{\lbrack
\psi ]}\right) (-\infty ,r]=\nu _{F_{\psi }^{g(Q)}}(-\infty ,r]$.

\item  Using $f(P)=\mathcal{F}^{-1}\cdot f(Q)\cdot \mathcal{F}$ we have $%
F_{\psi }^{f(P)}=\sum_{\left\| \psi _{s}\right\| \neq 0}F_{\psi
_{s}}^{f(P)}=\sum_{\left\| \psi _{s}\right\| \neq 0}F_{\mathcal{F}\psi
_{s}}^{f(Q)}$. Moreover $\left[ f(p)_{\ast }\mu _{\lbrack \psi ]}\right]
(-\infty ,r]=\sum_{\left\| \psi _{s}\right\| \neq 0}\iint_{\mathcal{\mathbb{R%
}}^{6}}\chi _{\left\{ f(p)\leq r\right\} }\cdot \frac{\left| \psi
_{s}\right| ^{2}\cdot \otimes \left| \mathcal{F}\psi _{s}\right| ^{2}}{%
\left\| \psi _{s}\right\| ^{2}}dqdp=$ $=\sum_{\left\| \psi _{s}\right\| \neq
0}\int_{\mathcal{\mathbb{R}}^{3}}\chi _{\left\{ f(p)\leq r\right\} }\cdot
\left| \mathcal{F}\psi _{s}\right| ^{2}\cdot dq=\left( \sum_{\left\| \psi
_{s}\right\| \neq 0}F_{\mathcal{F}\psi _{s}}^{f(Q)}\right) (r)=\nu _{F_{\psi
}^{f(P)}}(-\infty ,r]$.

\item  Applying the Thm. 7.17 of \textbf{[W]} to the operator $S$ it is not
difficult to prove that $\left( E_{(-\infty ,r]}^{S}\psi \right) (x,s)=\chi
_{(-\infty ,r]}(s)\cdot \psi _{s}(x)$ and $F_{\psi }^{S}(r)=\sum_{s\leq
r}\left\| \psi _{s}\right\| ^{2}$. Therefore $s_{\ast }\mu _{\lbrack \psi
]}=\nu _{F_{\psi }^{S}}$.
\end{enumerate}
\end{proof}

\begin{remark}
So for a suitable choice of a complex of barriers the function $%
g(q_{1},q_{2},q_{3})$ emulates the values of the operator $%
g(Q_{1},Q_{2},Q_{3})$. Then taken as $g$ the projection on the component $%
q_{j}$ there exists a quantum observable function giving the $j-th$
component of the position for $j=1,2,3$. If you have the skill to produce
concrete apparatuses with assigned barriers the components of the position
are outcomes for the label $(q,p)$! Analogously for the components of the
momentum (with different barriers).
\end{remark}

\begin{remark}
Therefore on the phase space each complex of barriers $\alpha $ with the map
$\widehat{(\cdot )}_{\alpha }:\Omega \rightarrow \mathcal{O}_{\alpha }$
(togeter with $\mu _{\cdot }$) defines a QMPS in the sense that $%
\left\langle \psi ,A\psi \right\rangle =\int_{Z}\widehat{A}_{\alpha \left[
\psi \right] }\cdot d\mu _{\left[ \psi \right] }$ (for each unitary vector $%
\psi $ and each self-adjoint operator $A$ such that both terms of the
equality are defined). The main differences with the usual QMPS is the
dependence of $\widehat{A}_{\alpha \left[ \psi \right] }$ on $\left[ \psi %
\right] $ and the lack of any explicit hypothesis of linearity on $A$ (cfr.
\textbf{[CZ], [Gr], [M], [P]}).
\end{remark}

\section{\textbf{\ Dynamics}}

\begin{definition}
For each $f$ in $\mathcal{O}_{\alpha b}$ the \textbf{quadratic function
associated to} $f$ is the function
\begin{equation*}
Q_{f}:\mathcal{H\diagdown }\left\{ 0\right\} \mathcal{\rightarrow }\mathbb{R}%
\text{ defined by }\mathcal{Q}_{f}(\psi )=\left\| \psi \right\| ^{2}\cdot
\int_{Z}f_{\left[ \psi \right] }(z)\cdot d\mu _{\left[ \psi \right] }
\end{equation*}
\end{definition}

\begin{remark}
The function $\mathcal{Q}_{f}$ can be considered defined and differentiable
on all $\mathcal{H}$ (even in $0$) since it is equal to $\left\langle \psi
,Op(f)\psi \right\rangle $.
\end{remark}

\begin{theorem}
For each $f$ in $\mathcal{O}_{\alpha b}$ we have:
\begin{equation*}
Op(f)\psi =\frac{1}{2}Grad_{\psi }\mathcal{Q}_{f}
\end{equation*}
\end{theorem}

\begin{proof}
Let $A=Op(f)$ for every $\varphi $ in $\mathcal{H}$ we have: $RE\left\langle
\varphi ,\frac{1}{2}Grad_{\psi }\mathcal{Q}_{f}\right\rangle =\frac{1}{2}RE%
\left[ \varphi _{\psi }(\mathcal{Q}_{f})\right] =\frac{1}{2}RE\left[
\lim_{t\rightarrow 0}\frac{1}{t}\left( \left\langle \psi +t\varphi ,A\psi
+tA\varphi \right\rangle -\left\langle \psi ,A\psi \right\rangle \right) %
\right] =$ $RE\left\langle \varphi ,A\psi \right\rangle $ therefore $A\psi =%
\frac{1}{2}Grad_{\psi }\mathcal{Q}_{f}$.
\end{proof}

\begin{definition}
Given $f$ and $g$ in $\mathcal{O}_{\alpha b}$ we define $\left\{ f,g\right\}
_{\alpha }=\widehat{\left( \left\{ Op(f),Op(g)\right\} \right) }_{\alpha }$
(where $\left\{ A,B\right\} =-\frac{i}{2}\left[ A,B\right] $) and $f\circ
_{\alpha }g=\widehat{\left( Op(f)\circ Op(g)\right) }_{\alpha }$ (where $%
A\circ B=\frac{1}{2}\left( AB+BA\right) $).
\end{definition}

\begin{remark}
Note that here: $A\cdot B=A\circ B+i\left\{ A,B\right\} $. The space $%
\mathcal{O}_{\alpha b}$ with $\left\{ \cdot ,\cdot \right\} _{\alpha }$ is a
Lie-algebra and $\mathcal{O}_{\alpha b}$ with $\circ _{\alpha }$ is a Jordan
algebra. In the complexification of $\mathcal{O}_{\alpha b}$ we can consider
the product $f\times _{\alpha }g=\widehat{\left( Op(f)\cdot Op(g)\right) }%
_{\alpha }=f\circ _{\alpha }g+i\cdot \left\{ f,g\right\} _{\alpha }$ that
makes it an associative algebra. We have: $\left\langle \left\langle \left\{
f,g\right\} _{\alpha }\right\rangle \right\rangle _{\psi }=\left\langle
\left\{ Op(f),Op(g)\right\} \right\rangle _{\psi }$, $\left\langle
\left\langle f\circ _{\alpha }g\right\rangle \right\rangle _{\psi
}=\left\langle Op(f)\circ Op(g)\right\rangle _{\psi }$ and $\left\langle
f\times _{\alpha }g\right\rangle _{\psi }=\left\langle Op(f)\cdot
Op(g)\right\rangle _{\psi }$. Moreover: $Q_{f\times _{\alpha }g}(\psi )=%
\frac{1}{4}\left\langle Grad_{\psi }Q_{f},Grad_{\psi }Q_{g}\right\rangle $, $%
Q_{f\circ _{\alpha }g}(\psi )=$ $RE\left[ Q_{f\times _{\alpha }g}(\psi )%
\right] $\ and $Q_{\left\{ f,g\right\} _{\alpha }}(\psi )=IM\left[
Q_{f\times _{\alpha }g}(\psi )\right] $.
\end{remark}

\begin{theorem}
Let $H$ be a bounded self-adjoint operator on $\mathcal{H}$ , let $\alpha
_{\cdot \text{ }}$be a complex of barriers, let $h=\widehat{H}_{\alpha }$
and let $\left\{ \psi _{t}\right\} _{t\in \mathbb{R}}$ be a differentiable
unitary path in $\mathcal{H}$.

Are equivalent:

\begin{enumerate}
\item  $i\frac{d}{dt}|_{t=t_{0}}\psi _{t}=H\psi _{t_{0}}$

\item  $\frac{d}{dt}|_{t=t_{0}}\int_{Z}f_{\left[ \psi _{t}\right] }\cdot
d\mu _{\left[ \psi _{t}\right] }=2\cdot \int_{Z}\left\{ f,h\right\} _{\alpha %
\left[ \psi _{t_{0}}\right] }\cdot d\mu _{_{\left[ \psi _{t_{0}}\right] }}$
for every $f$ in $\mathcal{O}_{\alpha b}$
\end{enumerate}
\end{theorem}

\begin{proof}
$\left[ 1)\Longrightarrow 2)\right] $ $\frac{d}{dt}|_{t=t_{0}}\int_{Z}f_{%
\left[ \psi _{t}\right] }\cdot d\mu _{\left[ \psi _{t}\right] }=\frac{d}{dt}%
|_{t=t_{0}}\left\langle \psi _{t},Op(f)\psi _{t}\right\rangle =\left\langle
\psi _{t_{0}},-i\left[ Op(f),H\right] \psi _{t_{0}}\right\rangle =$ $%
=\int_{Z}\widehat{\left( -i\left[ Op(f),H\right] \right) }_{\alpha _{\left[
\psi _{t_{0}}\right] }}\cdot d\mu _{_{\left[ \psi _{t_{0}}\right]
}}=2\int_{Z}\left\{ f.h\right\} _{\alpha \left[ \psi _{t_{0}}\right] }\cdot
d\mu _{_{\left[ \psi _{t_{0}}\right] }}$.

$\left[ 2)\Longrightarrow 1)\right] $ Let's fix $t_{0}$ in $\mathbb{R}$ and
a bounded self-adjoint operator $A$ such that $A\psi _{t_{0}}=\frac{d}{dt}%
|_{t=t_{0}}\psi _{t}+iH\psi _{t_{0}}$. Then taken $f=\widehat{A}_{\alpha }$:
$2\left\| \frac{d}{dt}|_{t=t_{0}}\psi _{t}+iH\psi _{t_{0}}\right\| ^{2}=%
\frac{d}{dt}|_{t=t_{0}}\left\langle \psi _{t},A\psi _{t}\right\rangle
-\left\langle \psi _{t_{0}},-i\left[ A,H\right] \psi _{t_{0}}\right\rangle =$

$=\frac{d}{dt}|_{t=t_{0}}\int_{Z}f_{\left[ \psi _{t}\right] }\cdot d\mu _{%
\left[ \psi _{t}\right] }-2\int_{Z}\left\{ f,h\right\} _{\alpha \left[ \psi
_{t_{0}}\right] }\cdot d\mu _{_{\left[ \psi _{t_{0}}\right] }}=0$.
\end{proof}

\begin{definition}
For every observable function $f$ and every $\left[ \psi \right] $ in $%
\mathbb{P(\mathcal{H})}$ the\textbf{\ dispersion of }$f$\textbf{\ on }$\left[
\psi \right] $ is the non-negative number $\Delta _{\psi }^{f}=\sqrt{%
\left\langle \left\langle \left[ f-\left\langle f\right\rangle _{\psi }%
\right] ^{2}\right\rangle \right\rangle _{\psi }}=\sqrt{\left\langle
\left\langle f^{2}\right\rangle \right\rangle _{\psi }-\left\langle
\left\langle f^{{}}\right\rangle \right\rangle _{\psi }^{2}}$.
\end{definition}

\begin{remark}
$\Delta _{\psi }^{f}=\Delta _{\psi }^{Op(f)}$ in fact: $\left( \Delta _{\psi
}^{f}\right) ^{2}=\left\langle \left\langle \left[ f-\left\langle
f\right\rangle _{\psi }\cdot 1\right] ^{2}\right\rangle \right\rangle _{\psi
}=\left\langle \left[ Op(f)-\left\langle f\right\rangle _{\psi }\cdot I%
\right] ^{2}\right\rangle _{\psi }=\left( \Delta _{\psi }^{Op(f)}\right)
^{2} $
\end{remark}

\begin{theorem}
(Heisenberg's uncertainty principle) For every $f$ and $g$ in $\mathcal{O}%
_{\alpha b}$ we have $\Delta _{\psi }^{f}\cdot \Delta _{\psi }^{g}\geq
\left| \left\langle \left\langle \left\{ f,g\right\} _{\alpha }\right\rangle
\right\rangle _{\psi }\right| $
\end{theorem}

\begin{proof}
$\Delta _{\psi }^{f}\cdot \Delta _{\psi }^{g}=\Delta _{\psi }^{Op(f)}\cdot
\Delta _{\psi }^{Op(g)}\geq \left| \left\langle \left\{ Op(f),Op(g)\right\}
_{\alpha }\right\rangle _{\psi }\right| =\left| \left\langle \left\langle
\left\{ f,g\right\} _{\alpha }\right\rangle \right\rangle _{\psi }\right| $
\end{proof}

\begin{definition}
A \textbf{complex of equivalences} is a family $\sigma =\left\{ \sigma
_{\lbrack \psi ]}\right\} _{[\psi ]\in \mathbb{P(\mathcal{H})}}$ of borel
equivalences $\sigma _{\lbrack \psi ]}:Z\rightarrow ]0,1[$ (defined out of
null borel subsets) with $\sigma _{\lbrack \psi ]\ast }\mu _{\lbrack \psi
]}=\lambda _{]0,1[}$ for each $[\psi ]$ in $\mathbb{P(\mathcal{H})}$.
\end{definition}

\begin{definition}
A unitary operator $\mathcal{U}$ together with a complex of equivalences $%
\sigma $ defines on the space of complete states an \textbf{automorphism }$%
\widehat{\mathcal{U}}_{\sigma }:\mathcal{C\rightarrow C}$ by:
\begin{equation*}
\widehat{\mathcal{U}}_{\sigma }([\psi ],\alpha _{\lbrack \psi ]},z)=\left( [%
\mathcal{U}\psi ],\left( \alpha _{\lbrack \psi ]}\circ \sigma _{\lbrack \psi
]}^{-1}\circ \sigma _{\lbrack \mathcal{U}\psi ]}\right) _{[\mathcal{U}\psi
]},\left( \sigma _{\lbrack \mathcal{U}\psi ]}^{-1}\circ \sigma _{\lbrack
\psi ]}\right) (z)\right)
\end{equation*}
\end{definition}

\begin{remark}
Given $\mathcal{U}$ and $\sigma $ the automorphism $\widehat{\mathcal{U}}%
_{\sigma }$ can be given equivalently by: $\widehat{\mathcal{U}}_{\sigma
}([\psi ],\alpha _{\lbrack \psi ]},z)=\left( [\mathcal{U}\psi ],\left(
\alpha _{\lbrack \psi ]}\circ \tau _{\lbrack \psi ]}^{-1}\right) _{[\mathcal{%
U}\psi ]},\tau _{\lbrack \psi ]}(z)\right) $ where $\tau _{\lbrack \psi
]}=\sigma _{\lbrack \mathcal{U}\psi ]}^{-1}\circ \sigma _{\lbrack \psi
]}:Z_{[\psi ]}\rightarrow Z_{[\mathcal{U}\psi ]}$ is a borel equivalence
(defined out of $\mu -$null borel subsets) with $\tau _{\lbrack \psi ]\ast
}\mu _{\lbrack \psi ]}=\mu _{\lbrack U\psi ]}$.
\end{remark}

\begin{remark}
Written $\widehat{\mathcal{U}}_{\sigma }([\psi ],\alpha ,z)=\left( [\mathcal{%
U}\psi ],\alpha ^{\prime },z^{\prime })\right) $ note that $\alpha ^{\prime
}(z^{\prime })=\alpha (z)$.
\end{remark}

\begin{remark}
Each automorphism $\widehat{\mathcal{U}}_{\sigma }$ induces the projective
transformation $\ \left[ \mathcal{U}\right] $ $\mathcal{:}\mathbb{P(\mathcal{%
H})\rightarrow P(\mathcal{H})}\mathcal{\ \ }$ defined by $\left[ \mathcal{U}%
\right] [\psi ]=$ $[\mathcal{U}\psi ]$. Fixed a complex of equivalences $%
\sigma $ we have $\widehat{\mathcal{U}}_{\sigma }=\widehat{\mathcal{V}}%
_{\sigma }$ if and only if $\left[ \mathcal{U}\right] =\left[ \mathcal{V}%
\right] $. The automorphisms with a fived $\sigma $ make a group isomorphic
to the group of projective transformations of $\mathbb{P(\mathcal{H})}$.
\end{remark}

\begin{theorem}
Given a 1-parameter group $\left\{ \mathcal{U}_{t}\right\} _{t\in \mathbb{R}%
} $ of unitary transformations of $\mathcal{H}$ and a complex of
equivalences $\sigma $, the family $\left\{ \widehat{\mathcal{U}}_{t\sigma
}\right\} _{t\in \mathbb{R}}$ is a 1-parameter group of automorphisms of $%
\mathcal{C}$ inducing the 1-parameter group $\left\{ \left[ \mathcal{U}_{t}%
\right] \right\} _{t\in \mathbb{R}}$ of projective transformations of $%
\mathbb{P(\mathcal{H})}$.
\end{theorem}

\begin{proof}
Obvious.

\begin{remark}
A 1-parameter group of automorphisms of $\mathcal{C}$ gives a deterministic
evolution for every initial complete state $([\psi _{0}],\alpha _{0[\psi
_{0}]},z_{0})$. Every evolution $\left\{ \mathcal{U}_{t}\psi _{0}\right\}
_{t\in \mathbb{R}}$ in $\mathcal{H}$ of a unitary vector $\psi _{0}$ with $%
\left\{ \mathcal{U}_{t}\right\} _{t\in \mathbb{R}}$ 1-parameter group of
unitary transformations, can be seen, in several ways, as the apparent part
of an evolution of complete states.

\begin{theorem}
For every self-adjoint operator $A$, every complex of equivalences $\sigma $
and every unitary operator $\mathcal{U}$ we have:
\begin{equation*}
\widehat{\mathcal{U}^{-1}A\mathcal{U}}=\widehat{A}\circ \widehat{\mathcal{U}}%
_{\sigma }
\end{equation*}
\end{theorem}
\end{remark}
\end{proof}

\begin{proof}
$\widehat{\left( \mathcal{U}^{-1}A\mathcal{U}\right) }([\psi ],\alpha
_{\lbrack \psi ]},z)=\left( F_{\mathcal{U}\psi }^{A}\right) ^{(-1)}(\alpha
_{\lbrack \psi ]}(z))$ and $\left( \widehat{A}\circ \widehat{\mathcal{U}}%
_{\sigma }\right) ([\psi ],\alpha _{\lbrack \psi ]},z)=$

$=\widehat{A}([\mathcal{U}\psi ],\left( \alpha _{\lbrack \psi ]}\circ \sigma
_{\lbrack \psi ]}^{-1}\circ \sigma _{\lbrack \mathcal{U}\psi ]}\right) _{[%
\mathcal{U}\psi ]},\left( \sigma _{\lbrack \mathcal{U}\psi ]}^{-1}\circ
\sigma _{\lbrack \psi ]}\right) (z))=$

$=\left( F_{\mathcal{U}\psi }^{A}\right) ^{(-1)}\left( \alpha _{\lbrack \psi
]}\circ \sigma _{\lbrack \psi ]}^{-1}\circ \sigma _{\lbrack \mathcal{U}\psi
]}\circ \sigma _{\lbrack \mathcal{U}\psi ]}^{-1}\circ \sigma _{\lbrack \psi
]}\right) =\left( F_{\mathcal{U}\psi }^{A}\right) ^{(-1)}(\alpha _{\lbrack
\psi ]}(z))$.
\end{proof}

\begin{theorem}
Given a bounded self-adjoint operator $A$, a 1-parameter group $\left\{
\mathcal{U}_{t}\right\} _{t\in \mathbb{R}}$ of unitary transformations of $%
\mathcal{H}$, a unitary vector $\psi _{0\text{ }}$, a complex of barriers $%
\alpha _{0}$ \ we have:
\begin{equation*}
\left\langle \psi _{t},A\psi _{t}\right\rangle =\int_{Z}\widehat{A}_{\alpha
_{0}}([\psi _{t}],z)\cdot d\mu _{_{\lbrack \psi _{0}]}}(z)
\end{equation*}
where $\psi _{t}=\mathcal{U}_{t}\psi _{0}$.
\end{theorem}

\begin{proof}
Fixed a complex of equivalences $\sigma $ we have: $\left\langle \psi
_{t},A\psi _{t}\right\rangle =\left\langle \psi _{0},\left( \mathcal{U}%
_{t}^{-1}A\mathcal{U}_{t}\right) \psi _{0}\right\rangle =\int_{Z}\widehat{%
\left( \mathcal{U}_{t}^{-1}A\mathcal{U}_{t}\right) }_{\alpha _{0}[\psi
_{0}]}(z)\cdot d\mu _{_{\lbrack \psi _{0}]}}=$ $=\int_{Z}\widehat{\left(
A\right) }(\widehat{\mathcal{U}}_{t\sigma })([\psi _{0}],\alpha _{0[\psi
_{0}]},z)\cdot d\mu _{_{\lbrack \psi _{0}]}}(z)=$ $=\int_{Z}\left( F_{\psi
_{t}}^{A}\right) ^{(-1)}\left( \alpha _{0[\psi _{0}]}\circ \sigma _{\lbrack
\psi _{0}]}^{-1}\circ \sigma _{\lbrack \psi _{0}]}\circ \sigma _{\lbrack
\psi _{t}]}^{-1}\circ \sigma _{\lbrack \psi _{t}]}(z)\right) \cdot d\mu
_{_{\lbrack \psi _{0}]}}(z)=$ $=\int_{Z}\widehat{\left( A\right) }_{\alpha
_{0}[\psi _{t}]}(z)\cdot d\mu _{_{\lbrack \psi _{0}]}}(z)$.
\end{proof}

\begin{remark}
The attended value of $A$ on $\psi _{t}$ (evolution of $\psi _{0\text{ }}$
at the time $t$) is equal to the mean value of $\widehat{A}_{\alpha _{0}}$
on the points $([\psi _{t}],z)$ (''evolutions'' of the points $([\psi
_{0}],z)$ at the time $t$).
\end{remark}

\section{\textbf{BIBLIOGRAPHY}}

\

\textbf{[CZ]\ }\ T.L. Curtright and C.K. Zachos: Quantum mechanics in phase
space.World Scientific - Singapore (2005)

\textbf{[G] }\ \ \ \ A. M. \ Gleason: Measures on the closed subspaces of a
Hilbert space. Jour. of Math. and Mech. Vol. 6 (1957) \ 885-89

\textbf{[Gr]} \ \ H.J. Groenewold: On the principles of elementary quantum
mechanics. Physica 12 (1946) 405-460

\textbf{[KS]} \ \ J.L. Kelley and T.P. Srinivasan: Measure and integral. \
Springer-Verlag - NY 1988

\textbf{[M] }\ \ \ J.E.Moyal: Quantum mechanics as a statistical theory. \
Proc. Cambridge Phil. Soc. 45 (1949) 99-124

\textbf{[P]} \ \ \ \ J. C. T. Pool: Mathematical aspects of the Weyl
correspondence. Jour. of Math. Phys. Vol. 7 N. 1 Jan. 1966

\textbf{[Ro] \ }C. Rovelli: Relational Quantum Mechanics. Int. Jour. of
Teor. Phys., 35 (1996) 1637-1678

\textbf{[Ry]} \ \ \ H. L. Royden: Real analysis - The Macmillan company -
London 1968

\textbf{[W]} \ \ J. Weidmann: Linear Operators in Hilbert Spaces.
Springer-Verlag, NY 1980

\end{document}